\documentclass[letter,printer]{aa}

\usepackage{graphicx}
\usepackage[position=top]{subfig}
\usepackage[final]{pdfpages}
\usepackage{amsmath}
\usepackage[utf8]{inputenc}
\usepackage{epsfig,amsmath,amssymb}
\usepackage[varg]{txfonts}

\def\ga{\,\hbox{\hbox{$ > $}\kern -0.8em \lower 1.0ex\hbox{$\sim$}}\,}
\def\la{\,\hbox{\hbox{$ < $}\kern -0.8em \lower 1.0ex\hbox{$\sim$}}\,}
\def\beq{\begin{equation}}
\def\eeq{\end{equation}}

\titlerunning{Magnetic fields of spiral galaxies}
\authorrunning{Ntormousi}

\begin{document}

%
\title{Magnetic fields in massive spirals: \\ The role of feedback and initial conditions}
\author{Evangelia Ntormousi\inst{1,2}}
\date{Received -- / Accepted --}

\institute{
Foundation for Research and Technology (FORTH), 
Nikolaou Plastira 100, Vassilika Vouton
GR - 711 10, Heraklion, Crete, Greece
\\
\and
Department of Physics and ITCP, University of Crete, 71003 Heraklion, Greece}

\abstract
{Magnetic fields play a very important role in the evolution of galaxies through their direct impact on star formation and stellar feedback-induced turbulence. However, their co-evolution with these processes has still not been thoroughly investigated, and the possible effect of the initial conditions is largely unknown.}
{This letter presents the first results from a series of high-resolution numerical models, aimed at deciphering the effect of the initial conditions and of stellar feedback on the evolution of the galactic magnetic field in isolated, Milky-Way-like galaxies.}
{The models start with an ordered, either poloidal or toroidal, magnetic field of varying strength, and are evolved with and without supernova feedback. They include a dark matter halo, a stellar and a gaseous disk, as well as the appropriate cooling and heating processes for the interstellar medium.}
{Independently of the initial conditions, the galaxies develop a turbulent velocity field and a random magnetic field component in under 15 Myrs. Supernova feedback is extremely efficient in building a random magnetic field component up to large galactic heights. However, a random magnetic field emerges even in runs without feedback, which points to an inherent instability of the ordered component.}
{Supernova feedback greatly affects the velocity field of the galaxy up to large galactic heights, and helps restructure the magnetic field up to 10 kpc above the disk, independently of the initial magnetic field morphology. On the other hand, the initial morphology of the magnetic field can accelerate the development of a random component at large heights. These effects have important implications for the study of the magnetic field evolution in galaxy simulations.}


\maketitle

\begin{table*}[!th]
\caption{Summary of the models. The magnetic field strength refers to the center of the galaxy.}
  \begin{tabular*}{\linewidth}{@{\extracolsep{\fill}}llll}
    \hline
    {\textbf{Model}}  &  {\textbf{Feedback}}  &  {\textbf{Magnetic field strength ($\mu$G)}}  & {\textbf{Magnetic field morphology}} \\ 
    \hline
    \hline
     M\_b100\_T & no & 0.1 & toroidal  \\
     M\_fb\_b100\_T & yes & 0.1 & toroidal  \\
     M\_b1\_T & no & 1 & toroidal   \\
     M\_fb\_b1\_T & yes & 1 & toroidal   \\
     M\_b100\_P & no & 0.1 & poloidal   \\
     M\_fb\_b100\_P & yes & 0.1 & poloidal  \\
    \hline
  \end{tabular*}
\label{sim_table}
\end{table*}

\section{Introduction}

Magnetic fields play a crucial role in all galactic processes: they can affect the evolution of turbulence, the flux of cosmic rays, the formation of the cold
molecular gas that is necessary for star formation, and the evolution of feedback regions. In essence, magnetic fields influence all the internal processes that redistribute mass, energy and momentum in a galaxy.

However, the observational methods available to measure them suffer from inherent uncertainties that prevent us from obtaining the full picture. The little we do know, mainly from Faraday rotation and synchrotron emission measurements, is that the magnetic field in galaxies like our own follows the large-scale spiral pattern, and that it appears to have an important turbulent component \citep{Beck_1996}.  From the recent, full-sky dust polarization studies with Planck, we also know that the magnetic field of the Milky Way is probably in rough equipartition with turbulence \citep{Soler_2013,Planck2015}.  
Clearly, the interpretation of these observations requires numerical models targeted at the co-evolution of the magnetic field and the galaxy.

So far the vast majority of galactic evolution models have ignored magnetic fields. However, the increasing realization that galaxy evolution cannot be fully understood without this essential component, together with the necessary computational advances, have led to an increasing number of direct, full galaxy simulations that include magnetization (e.g. \cite{Su2017,Khoperskov17}).  Much of the effort in this field is targeted at the problem of cosmic magnetic field amplification. For instance, \citet{Beck_2012} studied the amplification of magnetic fields in galactic halos by indirectly modeling stellar feedback. \citet{Pakmor_Springel_2013} simulated weakly magnetized disk galaxies with a moving mesh code, and found that stellar feedback led to a rapid amplification of the magnetic field, that brought it in pressure equilibrium with the gas.  \citet{Rieder_Teyssier_2016,Rieder_Teyssier_2017} simulated cosmological dwarf galaxies, showing a significant contribution of feedback-induced turbulence to driving a dynamo, as well as a galactic wind that re-distributed the field into the intergalactic medium.  Clearly, stellar feedback and magnetic fields largely regulate each other through turbulence.  
Nonetheless, simulations aiming at long-term evolution studies of the galaxy cannot capture the full dynamical range of turbulence, or explore different initial magnetic field configurations.

This letter presents a series of high-resolution, non-cosmological numerical models, including gas cooling, star formation, and supernova feedback, targeted at exploring the role of the initial magnetic field configuration and of stellar feedback in the evolution of the magnetic field of a massive spiral like the Milky Way. These models can reach the small scales where star formation and feedback act, with enough dynamical range to follow the effects of turbulence on the global magnetic field. 

The numerical method is explained in the Section \ref{numerics}, the results are presented in Section \ref{sec:results}, and discussed in Section \ref{sec:discussion}.

\begin{figure*}[h!]
   \centering
     \subfloat[M\_b100\_T]{
       \includegraphics[width=0.45\linewidth]{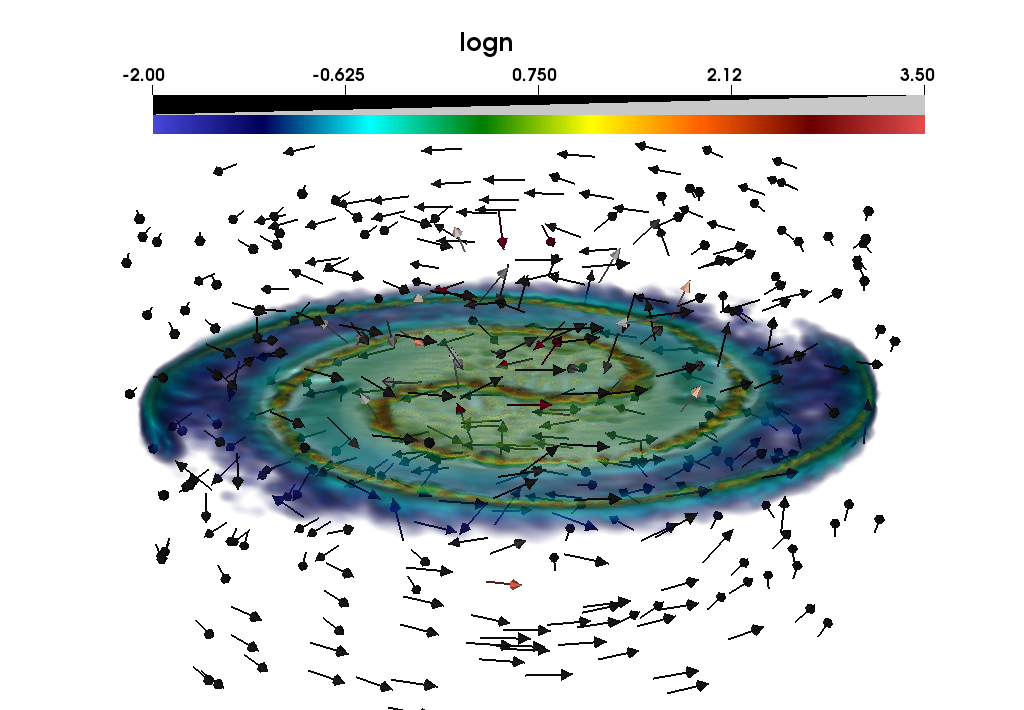}
     }
     \subfloat[M\_fb\_b100\_T]{
          \includegraphics[width=0.45\linewidth]{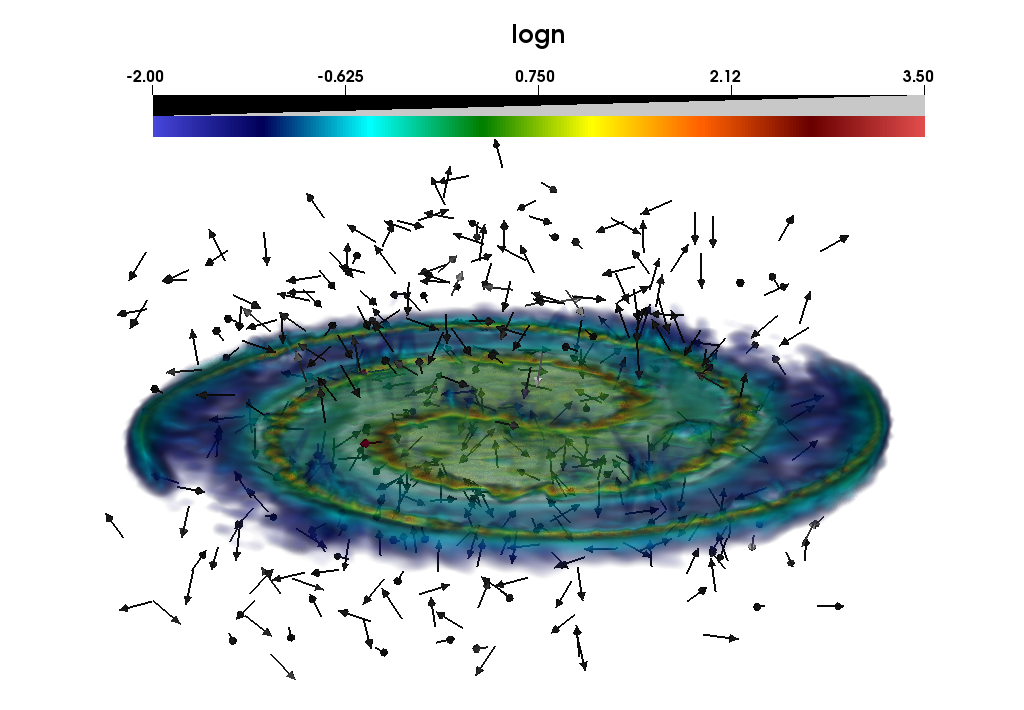}
     } \\ \vspace{-0.2cm}
   \subfloat[M\_b1\_T]{
       \includegraphics[width=0.45\linewidth]{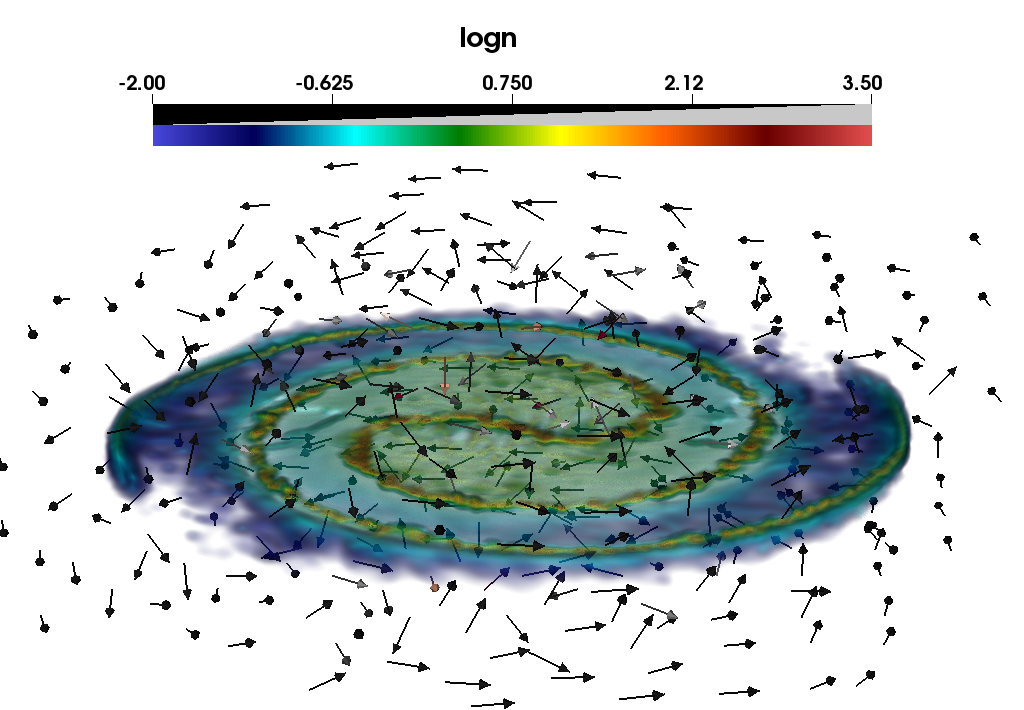}
     } 
   \subfloat[M\_fb\_b1\_T]{
       \includegraphics[width=0.45\linewidth]{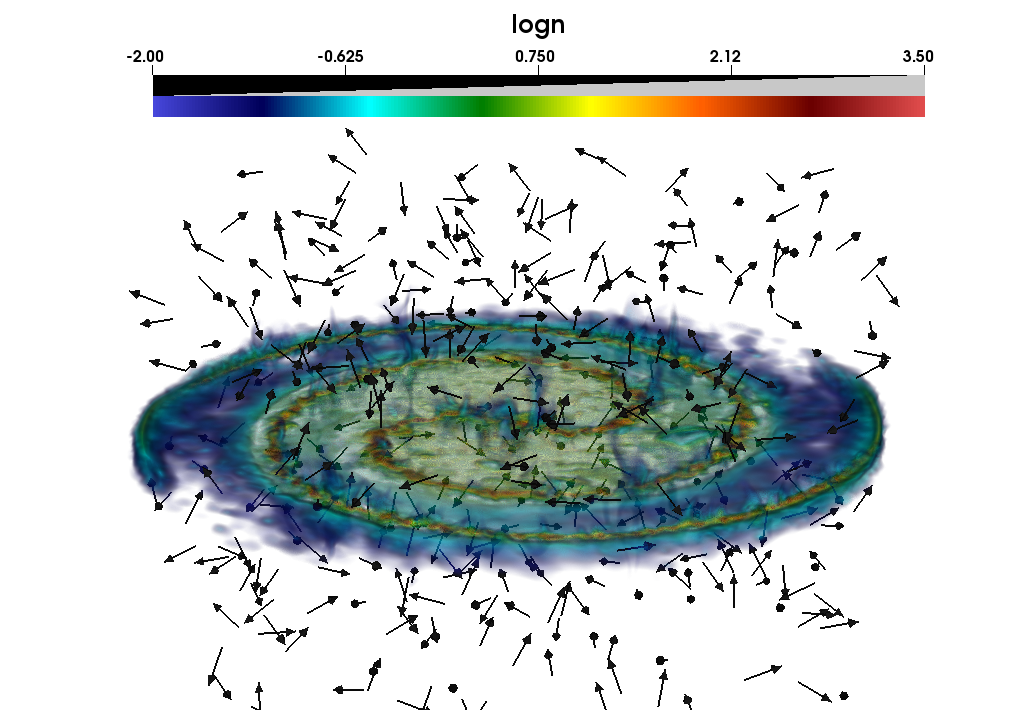}
    } \\  \vspace{-0.2cm}
  \subfloat[M\_b100\_P]{
       \includegraphics[width=0.45\linewidth]{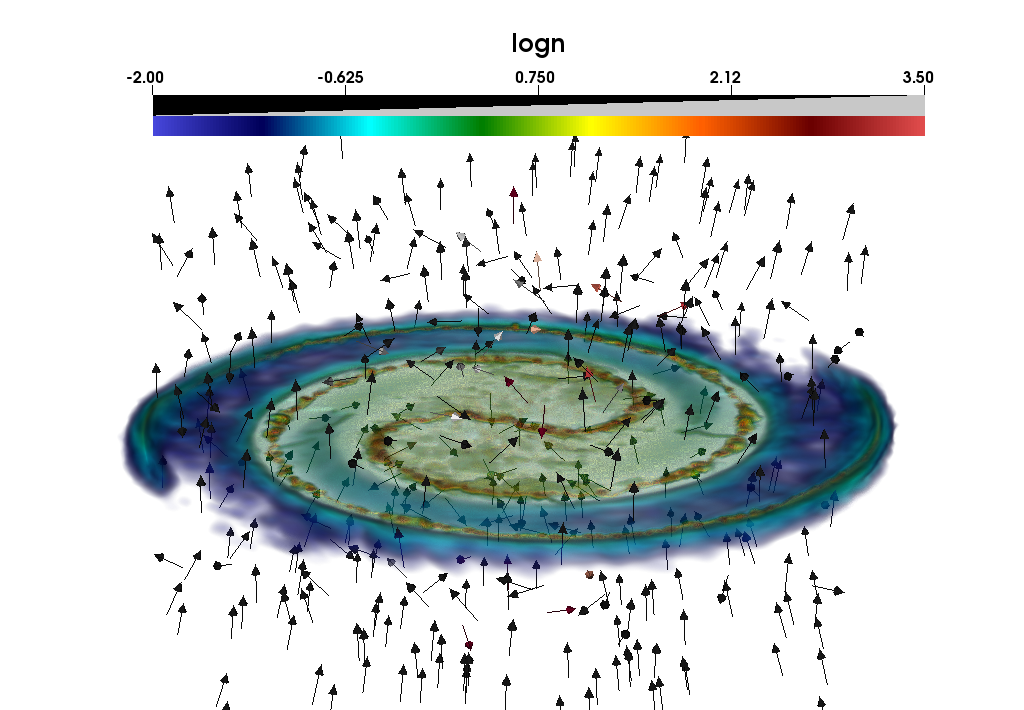}
    }
  \subfloat[M\_fb\_b100\_P]{
       \includegraphics[width=0.45\linewidth]{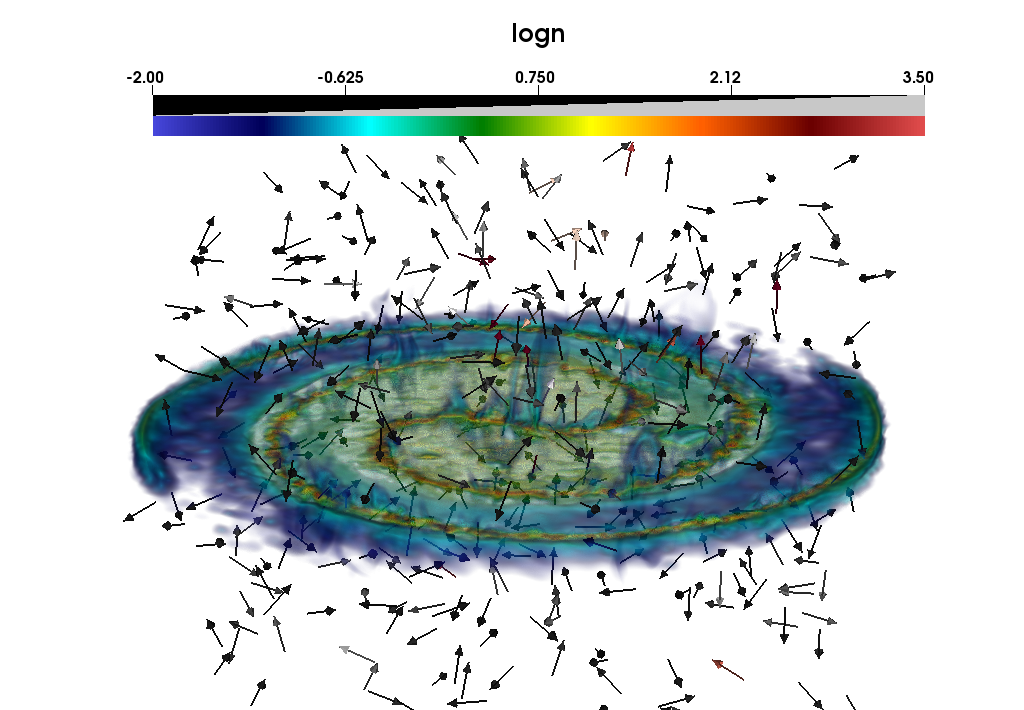}
    }
   \caption{The disk density in volume rendering, and the global magnetic field configuration in vectors after 15 Myrs of evolution.  The box size in this figure is 24 kpc.}
     \label{magnetic_evolution}
\end{figure*}
%
\begin{figure*}[h]
   \centering
   \subfloat[Magnetic field with time]{
          \includegraphics[width=0.3\linewidth]{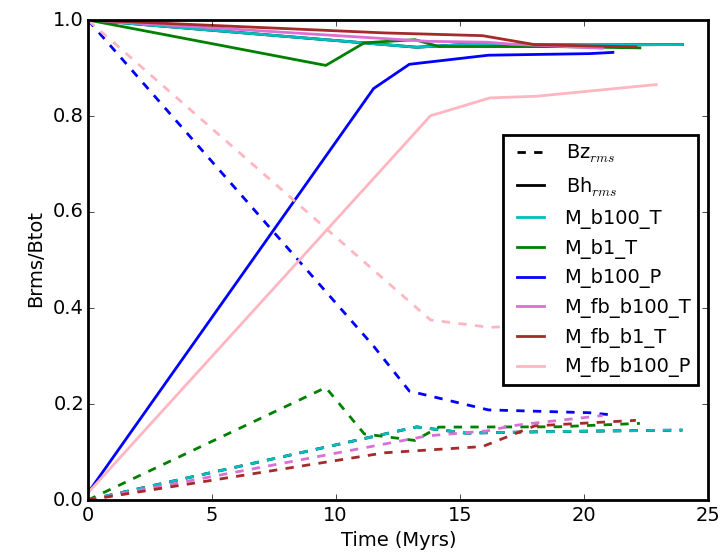}
          }
     \subfloat[Kinetic energy power spectra]{
       \includegraphics[width=0.3\linewidth]{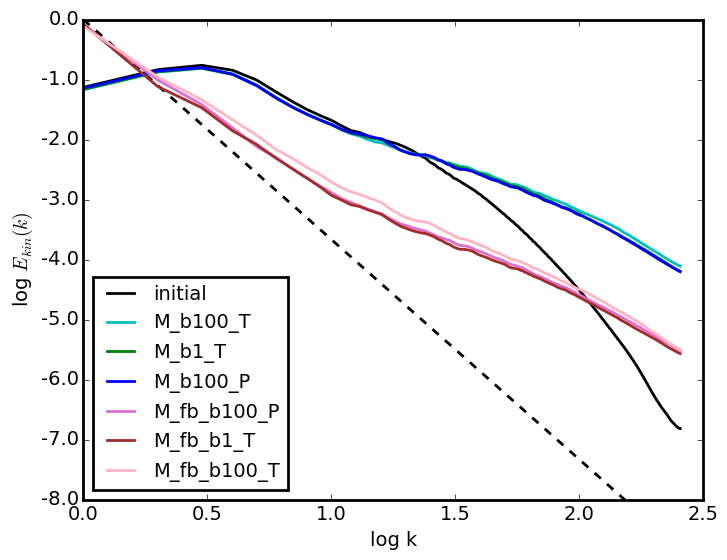}
     }
     \subfloat[Magnetic field power spectra]{
          \includegraphics[width=0.3\linewidth]{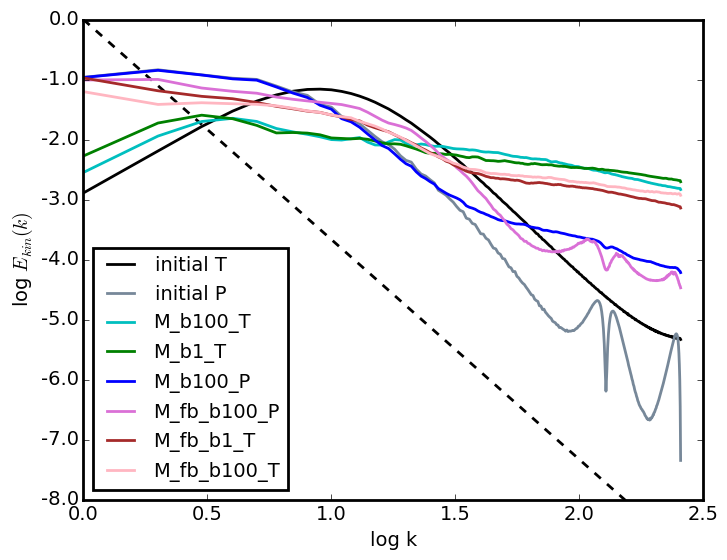}
     } 
   \caption{(a) Relative strengths of the total vertical and horizontal rms magnetic fields with time, (b) kinetic energy power spectra and (c) magnetic field power spectra. The black dashed lines in panels (b) and (c) show the Kolmogorov $\propto k^{-11/3}$ dependence.}
     \label{power_spectra}
\end{figure*}
%
\section{Method and setup}
\label{numerics}

\subsection{Numerical code}

The simulations are performed with the publicly available MHD code RAMSES \citep{Teyssier_02}, which solves the MHD equations on a Cartesian grid and has Adaptive Mesh Refinement (AMR) capabilities.

The equations solved by the code are:
\begin{eqnarray}
\frac{\partial\rho}{\partial t} +\nabla(\rho\bf v) = 0 \label{continuity} \\
\frac{\partial\bf{v}}{\partial t} + (\bf{v\cdot\nabla})\cdot\bf{v} + \frac{1}{\rho}\nabla P = -\nabla\phi \label{momentum} \\
\frac{\partial E_{tot}}{\partial t} +\nabla(~(E_{tot}+P)\bf{v} -(\bf{v}\cdot\bf{B})\cdot\bf{B} = -v\cdot\nabla\phi \label{energy} \\
\frac{\partial\bf{B}}{\partial t} -\nabla\times(\bf{v}\times\bf{B}) = 0 \label{induction} \\
\nabla\cdot\bf{B} = 0 
\end{eqnarray}
where $\rho$ the gas density, ${\bf{v}}$ the velocity, $E_{tot}$ the total energy, $P$ the pressure, 
${\bf{B}}$ the magnetic field and $\phi$ the gravitational potential. 
RAMSES uses a constrained transport scheme to evolve the magnetic field, which guarantees $\nabla\cdot{\bf{B}}=0$ always \citep{Fromang_2006}.  
This is a significant advantage with respect to codes that have to depend on divergence-cleaning algorithms, known to create spurious effects in studies of turbulent environments \citep{Balsara_2004}.  
\subsection{Initial setup}

The initial conditions are created using the DICE code \citep{Perret_2014,Perret_2016}. DICE is a MCMC-based initial condition generator that ensures the hydrostatic equilibrium of the galaxy.  Although DICE already comes with a RAMSES patch, it only allows a toroidal magnetic field configuration. More options for the magnetic field topology were added for the purposes of this work, which are open for download \footnote{\url{https://bitbucket.org/entorm/ramses_dice_magnetic}}.

We use a configuration of stellar and dark matter particles, as well as a hydrodynamical fluid, to simulate a Milky-Way-like galaxy (total mass $M_{tot}=2\cdot10^{12}~M_{\odot}$)  at redshift $z=0$ with different initial morphologies and strengths of the magnetic field. The Virial velocity of the galaxy is 200 km/sec, the mass fraction in stars is about $4.5\%$ (including a stellar bulge with a mass fraction of 0.5$\%$) and the mass fraction in gas is about $1\%$.  
The dark matter halo follows an NFW profile \citep{NFW96}, while the gas and stars are initially placed in exponential disks with a scale length of 9~kpc.  The galactic disk is perturbed with an $m=2$ perturbation, starting at 2~kpc and ending at 12kpc, which leads to the formation of a bar and two spiral arms. The mean gas temperature is set at $10^4$ K and subsonic turbulence, with an rms value of 8~km/sec is introduced throughout the disk. The disk gas has solar metallicity, while the halo gas has 10$^{-4}$ solar, both constant.   Appropriate cooling and heating rates for the interstellar medium are calculated according to \citet{Sutherland_Dopita_1993}.

Table \ref{sim_table} contains the initial parameters of the different galaxy models. The parameters varied here are the strength and the initial topology of the magnetic field.
The simulations start with either a toroidal field (model name ends with T), with a scale height and scale length of 1~kpc, or with a poloidal magnetic field  (model name ends with P), with a scale height of 1~kpc and a scale length of 2~kpc.  The poloidal magnetic field is actually model C from \citet{Ferriere_Terral_2014}.  
The magnetic field of the disk in runs M\_b1\_T and M\_fb\_b1\_T is initially just below equipartition with the kinetic energy of the gas in the center and in the interarm regions, and drops to a factor of 100 below equipartition in the spiral arms. All the other runs start off significantly below equipartition everywhere. In all models, the thermal energy is initially more than 100 times below the total kinetic energy of the gas, but in equipartition with the turbulent kinetic energy.
Unfortunately, the limit of initially global energy equipartition has not been explored in this work due to a lack of computational resources.

Star formation is simulated in all models by forming sink particles when the density exceeds 1000 cm$^{-3}$. Models whose name contains "fb" additionally include stellar feedback from supernovae, resulting from previously formed sink particles with a time delay of 3~Myrs.  Supernovae are implemented by injecting thermal energy into the cells around the sink particle according to the number of supernovae estimated for the predicted size of the formed stellar cluster.

AMR is used here to capture the complex dynamics of the disk.  In the box of 60~kpc, we use a coarse resolution of $256^3$ with four levels of refinement, reaching an effective $1024^3$ in the entire disk, and $2048^3$ in regions with active star formation. 

%
\section{Model evolution}
\label{sec:results}

Figure \ref{magnetic_evolution} shows the morphology of the galaxies and of their global magnetic field after 15 Myrs.  Independently of the initial conditions, there is a marked difference in the degree of turbulence in the magnetic field between the models with feedback and those without.  The models with feedback immediately develop a visible random component up to tens of kpc above the disk, while at the same heights the models without feedback largely retain the initial ordered morphology.  This seems to be independent of the initial magnetic field strength.

In particular, all models with an initially toroidal magnetic field develop a vertical component, while both models with an initially poloidal magnetic field develop a horizontal component (panel (a) of Fig. (\ref{power_spectra})). Interestingly, 22~Myrs into the evolution, all models tend to roughly the same component separation: about 10 per cent of the total rms magnetic field is in the vertical component, independently of the initial condition, or the presence of feedback.

The normalized kinetic energy power spectra, and volume-weighted magnetic field power spectra are shown in panels (b) and (c) of Fig. (\ref{power_spectra}) at 22~Myrs, for the full simulation box (L=60~kpc), smoothed at a 512$^3$ resolution. The black and grey curves show the initial toroidal and poloidal magnetic power spectra, respectively, while the dashed black line in both figures shows the Kolmogorov $P\propto k^{-11/3}$ dependence.

The kinetic energy power spectra are clearly divided into two families: the models with feedback, showing a Kolmogorov-like dependence on large scales, and the models without feedback, which remain close to the initial conditions on large scales. The fact that supernovae create power so quickly even in the halo implies a rapid communication between distant scales, due to the very high speed of sound in the halo ($c_s\simeq300~km/sec$), that permits a disturbance to travel at heights above 8kpc in 22~Myrs. On the other hand, there seems to be no influence of either the strength or the morphology of the initial magnetic field on the kinetic energy power spectra, which is consistent with the fact that the magnetic field stays significantly below equipartition for the entire length of these simulations.

On the contrary, the magnetic field power spectra keep a memory of their initial condition after one feedback cycle. Although all models develop a random magnetic field, as evidenced by the flat power spectra on scales smaller than a kpc, the ones with a poloidal magnetic field remain closer to the initial conditions on large scales, independently of feedback.  This is not surprising, since their initial magnetic field has a dominant vertical component, is therefore aligned with the feedback-induced outflow into the halo, and, in these timescales, is only deformed significantly in the disk. 
As in the kinetic energy power spectra, we observe again a clear separation into models with and without feedback at the largest scales, due to a rapid communication between scales.

\section{Discussion and conclusions}
\label{sec:discussion}

This letter presented the first results from high-resolution simulations of magnetized massive spirals, exploring the short-term evolution of their velocity and magnetic field under different physical conditions.

The most striking result is that, all galaxies, independently of the initial magnetic field morphology, quickly develop a random magnetic field component. A vertical component emerges in runs with an initially purely toroidal field, and a horizontal component in runs with an initially poloidal field. At 22 Myrs, all models seem to tend to the same relative rms component separation: about 10 percent vertical, 90 percent horizontal.  Although supernova feedback is very efficient in accelerating the process of redistributing magnetic energy and randomizing the field up to larger heights (a process reminiscent of the feedback "action at a distance" observed in star formation simulations \citep{Offner}), the new components appear also in runs without feedback.

This may be an indication of a Tayler-like instability of the galactic magnetic field, analogous to that of a stellar poloidal or toroidal field \citep{Bonnano_DS_2012, Bonnano_2008, Tayler_1973}, as a result of the axisymmetric (spiral arms) and turbulent disturbances in the density and velocity field. 
In this case, the instability seems to be growing much more slowly than the field redistribution due to feedback. Certainly such an investigation is of great interest for a follow-up paper.

Finally, the power spectra of kinetic energy and magnetic field show that:
\begin{enumerate}
\item The velocity field is very rapidly (<15~Myrs) affected by supernova feedback, which builds Kolmogorov-like turbulence up to tens of kiloparsecs above the disk due to the high temperatures of the halo gas.
\item The magnetic field is random at small scales, shaped by the inherent instability of the initial conditions and the feedback, whenever present. In fact, our magnetic power spectra are very different than those reported for example by \citet{Rieder_Teyssier_2017}, who found a Kolmogorov-like behavior of the magnetic field in small scales in their dynamo studies. It is possible that a longer evolution of the models could create a small-scale turbulent spectrum, but this cannot be deduced by the current state of the models. 
\end{enumerate}

\emph{Acknowledgments}

The author is grateful to Fabio Del Sordo, Kostantinos Tassis, Vassiliki Pavlidou, and to the anonymous referee for useful comments, as well as to 
Francesca Fragkoudi for advice on using DICE.
This research is funded by a Marie Curie Action of the European Union (Grant agreement number 749073).
Initial testing of the numerical setup was performed on the HPC resources of CINES, under an allocation led by Patrick Hennebelle. 

\bibliographystyle{apj}
\bibliography{spirals}
\label{lastpage}

\end{document}